\documentclass[12pt]{iopart}
\usepackage{graphicx}
\usepackage{cite}
\usepackage{xcolor}

\begin{document}
\title[Convolutional neural network for retrieval...]{Convolutional neural network for retrieval of the time-dependent bond length in a molecule from photoelectron momentum distributions}

\author{N I Shvetsov-Shilovski and M Lein}

\address{Institut f\"{u}r Theoretische Physik, Leibniz Universit\"{a}t Hannover, 30167 Hannover, Germany}

\ead{n79@narod.ru}

\begin{abstract}
We apply deep learning for retrieval of the time-dependent bond length in the dissociating two-dimensional H$_2^{+}$ molecule using photoelectron momentum distributions. We consider a pump-probe scheme and calculate electron momentum distributions from strong-field ionization by treating the motion of the nuclei classically, semiclassically or quantum mechanically. A convolutional neural network trained on momentum distributions obtained at fixed internuclear distances retrieves the time-varying bond length with an absolute error of $0.2$-$0.3$~a.u.
\end{abstract}


\submitto{\jpb}
\maketitle

\section{Introduction}

Development of techniques for static and dynamic molecular imaging is all-important for chemistry, biology, and material science. Strong-field physics  that studies the interaction of strong laser pulses with atoms and molecules (Refs.~\cite{Graefe2016,Lin2018}) offers new approaches for molecular imaging \cite{Agostini2016}. In particular, it was shown that momentum distributions of photoelectrons produced by a strong laser pulse encode spatiotemporal information about the parent ion. Laser-induced electron diffraction (LIED) and strong-field photoelectron holography (SFPH) are two methods of molecular imaging based on analysis of electron momentum distributions. In recent years, both of these methods have been intensively studied, see Ref.~\cite{Giovannini2023} for a review of LIED and, e.g., Refs.~\cite{Walt2017,Keil2017,Maurer2018,Tan2019,Brennecke2019,He2018a,He2018b} for new developments in SFPH.   

Electrons generated in strong-field ionization that do not return to their parent ions are often referred to as direct electrons. Typically they have low energies. Besides the direct electrons, there are also electrons that are driven back by the oscillating field of the laser pulse to the ions and scatter from them. For large scattering angles, these rescattered electrons create the high-energy plateau in the photoelectron energy spectrum. In LIED, both static and time-resolved information is extracted from momentum distributions of rescattered electrons. On the contrary, SFPH analyses holographic patterns that emerge in the low-energy part of photoelectron momentum distributions due to the interference of direct and rescattered electrons. While the LIED method employing high-energy rescattered electrons provides a way to probe the interior of a molecule on short distances, the SFPH technique is more sensitive to the molecular potential at large distances, as it uses the low-energy direct electrons. This leads to the idea that the advantages of both approaches should be combined. Even though LIED and SFPH were simultaneously implemented in experiment \cite{Walt2017}, the unification of these two methods is a difficult task that has not been solved yet.

The first problem for integration of LIED and SFPH is that these methods use different parts of PMDs. Second, the theoretical models that are used to interpret the LIED and SFPH experiments are different. The LIED usually employs the combination of quantitative rescattering theory \cite{Chen2009,Lin2010}, the independent-atom model \cite{Schafer1976,McCaffrey2008}, and the three-step model \cite{Krause1992,Corkum1993}. In contrast to this, the interference patterns of SFPH are analysed by using the direct numerical solution of the time-dependent Schr\"{o}dinger equation (TDSE) in single-active-electron approximation \cite{Huismans2011,Marchenko2011,Bian2011,Hickstein2012,Haertelt2016}, the strong-field approximation (see Refs.~\cite{Keldysh1964,Faisal1973,Reiss1980}) with rescattering \cite{Huismans2011,Huismans2012}, or the three-step model with interference \cite{Bian2011,Bian2012,Bian2014,Li2015}. Furthermore, all these models imply a number of approximations. As a result, the obtained information about a molecule may be not accurate enough. 

These issues can be addressed by application of machine learning (ML) which allows us to find complex mappings depending on many variables, often hidden in a large amount of data. ML has been used previously in strong-field physics. Among its applications in this field are the reconstruction of the intensity and the carrier-envelope phase (CEP) of short laser pulses from two-dimensional (2D) images (frequency-resolved optical gating traces \cite{Zahavy2018} and dispersion scan traces \cite{Kleinert2019}), the prediction of the high-order harmonic spectra for model di- and triatomic molecules \cite{Lytova2015}, and the efficient implementation of the trajectory-based Coulomb-corrected strong-field approximation (TCSFA) \cite{Yang2020} (see Refs.~\cite{Yan2010,Yan2012} for the formulation of the TCSFA). Recently, ML was also applied for creating movies of attosecond charge migration based on high-harmonic spectroscopy, the reconstruction of the geometrical structure of molecules in LIED \cite{Liu2021}, and the retrieval of the internuclear distance in a molecule from the PMD \cite{Shvetsov2022,Shvetsov2023}.

The convolutional neural network (CNN) trained in Ref.~\cite{Shvetsov2022} using the whole PMD can retrieve internuclear distance in a two-dimensional H$_2^{+}$ molecule with fixed nuclei with an absolute error less than $0.1$~a.u. The approach of Refs.~\cite{Shvetsov2022,Shvetsov2023} differs from the method used in Ref.~\cite{Liu2021}, where 2D differential cross sections were used to train a CNN aimed at the reconstruction of molecular structure. By training the CNN on distributions calculated by the direct solution of the TDSE, the studies of Refs.~\cite{Shvetsov2022,Shvetsov2023} avoid a number of approximations that are usually made in the LIED and SFPH methods. It is therefore of interest to apply the CNN \cite{Shvetsov2022,Shvetsov2023} for retrieval of the time-dependent internuclear distance in the case of moving nuclei.

In this paper we address the above-formulated problem. We use three different approaches to account for the motion of atomic nuclei, and we show that in all the three cases the CNN \cite{Shvetsov2022} trained on PMDs obtained for fixed internuclear distances retrieves the time-dependent bond length with good accuracy. The paper is organized as follows. In Sec.~II we review the architecture and training of the CNN as well as the method used for the solution of the TDSE. In Sec.~III we apply the CNN \cite{Shvetsov2022,Shvetsov2023} to momentum distributions produced by strong-field ionization of the dissociating H$_2^{+}$ molecule. The conclusions are given in Sec.~IV. Atomic units are used throughout the paper unless indicated otherwise.   

\section{Method}

A detailed description of the neural network used for retrieval of the bond length in the case of fixed nuclei, as well as of the method for the solution of the 2D TDSE are presented in Refs.~\cite{Shvetsov2022,Shvetsov2023}. Here we only briefly repeat the main points. The images used by the CNN as input are preprocessed by calculating the decimal logarithm of the normalized momentum distribution, i.e., $W=\log_{10}\left(\textrm{PMD}/\textrm{PMD}_{\textrm{\footnotesize max}}\right)$, and by setting $W=-5$ for all values smaller than $-5.$ Here PMD$_{\textrm{\footnotesize max}}$ is maximum of the distribution. By using bicubic interpolation we downsize the rectangular part of the image containing all values of W that exceed $-5$ to the size of $256 \times 128$ pixels. After rescaling of all the elements of the resulting matrix, such that the minimum value corresponds to $0$ and the maximum one to $255$, the images are given to the neural network. 

The CNN, which is implemented using the MATLAB package \cite{Matlab}, consists of five consecutive pairs of nonreducing convolutional layers and reducing average pooling layers. Each of the convolutional layers consists of $32$ filters with sizes $3\times3$ pixels. The last average pooling layer is connected to the dropout layer that randomly sets its input elements to zero with a given probability (20 \% in our case). This forces the CNN to develop a variety of ways to achieve the same result, and thus to prevent overfitting. The output of the dropout layer is received by the fully connected layer that predicts the internuclear distance $R$. We split the set of images into training and test sets in the ratio $0.75:0.25$. We use the mean squared error as the loss function, i.e., a measure of the deviation between the predictions of the CNN and true values of bond lengths over the training set. The neural network is trained by using the stochastic gradient descent optimizer. We begin the training with the learning rate of $10^{-3}$, which is then decreased by a factor of $10$ after $20$ training epochs. The loss function converges after about $30$ training epochs. 

We define the laser pulse in terms of the vector potential
\begin{equation}
\vec{A}\left(t\right)=\left(-1\right)^{n_p}\frac{F_0}{\omega}\sin^2\left(\frac{\omega t}{2n_p}\right)\sin\left(\omega t+\varphi\right)\vec{e}_{x},
\end{equation}
where $F_0$ is the field strength, $\omega$ is the laser frequency, $n_p$ is the number of optical cycles within the pulse, which lasts from $t=0$ to $t=\left(2\pi/\omega\right)\cdot n_{p}$, $\varphi$ is the CEP, and $\vec{e}_{x}$ is the unit vector in the polarization direction ($x$ axis). The electric field is $\vec{F}=-d\vec{A}/dt$. 

For the 2D H$_2^{+}$ molecular ion interacting with the laser pulse, the TDSE in the velocity gauge reads as
\begin{eqnarray}
&i\frac{\partial}{\partial t}\Psi\left(x,y,t\right)\nonumber \\
&=\left\{-\frac{1}{2}\left(\frac{\partial^2}{\partial x^2}+\frac{\partial^2}{\partial y^2}\right)-iA_{x}\left(t\right)\frac{\partial}{\partial x}+V\left(x,y\right)\right\}\Psi\left(x,y,t\right),
\label{tdse_f}
\end{eqnarray}
where $\Psi\left(x,y,t\right)$ is the wave function and 
\begin{equation}
V\left(x,y\right)=-\frac{1}{\sqrt{\left(x-\frac{1}{2}R\right)^2+y^2+a}}-\frac{1}{\sqrt{\left(x+\frac{1}{2}R\right)^2+y^2+a}}
\end{equation}
is the soft-core binding potential in the approximation of frozen nuclei. Here $R$ is the internuclear distance, and $a=0.64$ is the soft-core parameter. 

In order to solve the TDSE (2), we use the Feit-Fleck-Steiger split operator method \cite{Feit1982}. The wave function of the ground state is obtained by using the imaginary-time propagation. We use a computational box that extends over $x\in\left[-400,400\right]$ and $y\in\left[-200,200\right]$~a.u. and has its center at $\left(x=0,y=0\right)$. Our grid spacings for both directions are equal: $\Delta x=\Delta y=0.1954$~a.u. We use the time step $0.0184$~a.u. and we propagate the TDSE (\ref{tdse_f}) from $t=0$ (beginning of the pulse) to $t=4t_f$.  

The absorbing boundaries that allow us to prevent unphysical reflections from the edge of the computational box are implemented by multiplication of the wave function at every time step by the mask:
\begin{equation}
\label{cases}
M\left(x,y\right)=\cases{1&for $r\leq r_b$\\
\exp\left[-\beta\left(r-r_b\right)^2\right]&for $r > r_b$\\}
\label{mask}
\end{equation}
where $\beta=10^{-4}$, $r_b=150$~a.u., and $r=\sqrt{x^2+y^2}$. The electron momentum distributions are obtained by using the mask method \cite{Lein2002,Tong2006}.   

\section{Results and discussion}

\begin{figure}[h]
\centering
\includegraphics[width=.80\textwidth]{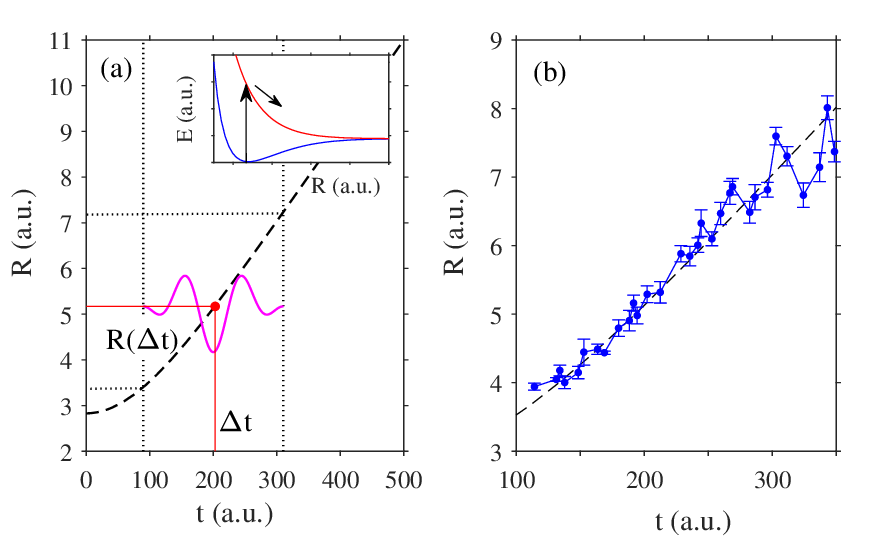}
\caption{(a) Retrieval of the time-dependent internuclear distance in the dissociating 2D H$_2^{+}$ molecule in a pump-probe scheme. The internuclear distance as a function of time after the excitation by the pump step (see inset) is shown by the black dashed curve. The two-cycle probe pulse (magenta curve) arriving after a certain time delay $\Delta t$ ionizes the molecule at a corresponding internuclear distance $R\left(\Delta t\right)$. The time delay refers to the center of the probe pulse. (b) Plot of averaged predictions of 5 CNNs for the internuclear distances (blue points) at different time delays compared to the time-dependent bond length (black dashed curve) obtained from the classical equation of motion.}
\label{calc_1}
\end{figure}

We analyse the retrieval of the time-dependent internuclear distance in the H$_{2}^{+}$ molecule by considering a pump-probe scheme. Nuclear motion is initiated by preparing the molecule in the first electronically excited state, which could be achieved by a suitable pump pulse. The molecule is ionized by a short and strong probe pulse acting after a certain time delay. This delay determines the internuclear distance at the ionization time, see Figure~\ref{calc_1}~(a). We use three different approaches in order to treat the nuclear motion. 

The first approach is based on the Born-Oppenheimer approximation ~\cite{Bransden1983}. More specifically, we assume that the nuclei move classically along the Born-Oppenheimer potential. Therefore, we find the internuclear distance $R\left(t\right)$ by integrating the Newton's equation of motion 
\begin{equation}
M\ddot{\vec{R}}=-\nabla_{\vec{R}}E_{e}\left(\vec{R}\right),
\label{Newton}
\end{equation}
where $M$ is the reduced mass of the nuclei, and $E_{e}\left(\vec{R}\right)$ is the excited state Born-Oppenheimer potential. We set $R\left(0\right)$ to the equilibrium internuclear distance in the ground state. The corresponding electron momentum distribution is obtained from the TDSE for an electron interacting with the probe pulse 
\begin{eqnarray}
& i\frac{\partial}{\partial t}\Psi\left(\vec{r},t\right)=\left[-\frac{1}{2}\left(\frac{\partial^2}{\partial x^2}+\frac{\partial^2}{\partial y^2}\right)-iA_{x}\left(t\right)\frac{\partial}{\partial x}\right.\nonumber \\
& \left.-\frac{1}{\sqrt{\left(x-R_{1}\left(t\right)\right)^2+y^2+a}}-\frac{1}{\sqrt{\left(x-R_{2}\left(t\right)\right)^2+y^2+a}}\right]\Psi\left(\vec{r},t\right).
\label{tdse_t}
\end{eqnarray}
Here $\vec{r}=x\vec{e}_{x}+y\vec{e}_{y}$, and $R_1\left(t\right)=R\left(t\right)/2$ and $R_2\left(t\right)=-R\left(t\right)/2$ are the positions of the nuclei. 

The CNNs \cite{Shvetsov2022,Shvetsov2023} were trained for ionization of the ground state of the H$_2^{+}$ molecule. In order to make them applicable to momentum distributions from ionization of the first excited electronic state, we use the transfer learning technique \cite{Goodfellow2016}. As in Ref.~\cite{Shvetsov2023}, we fix the first three convolutional layers and we use the learning rate $10^{-2}$ for the retraining. A set of $N=1000$ PMDs is needed to achieve the mean absolute error (MAE) $0.04$~a.u for the internuclear distance.

\begin{figure}[h]
\centering
\includegraphics[trim={0 0 0 -10}, width=.60\textwidth]{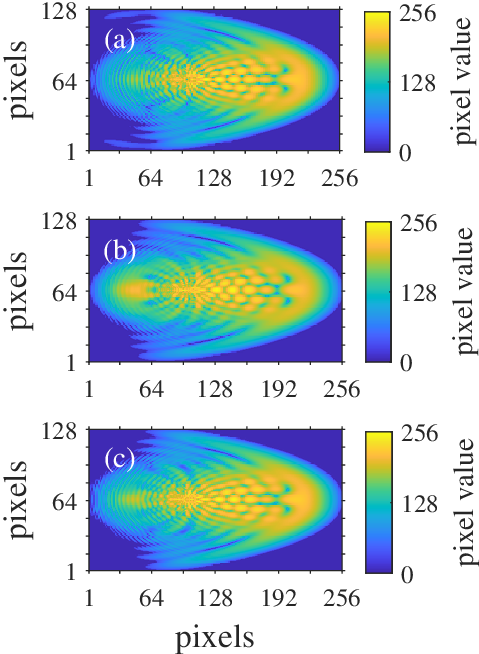}
\caption{Electron momentum distributions for ionization of the H$_{2}^{+}$ molecule for the internuclear distance $R=4.48$~a.u. as they are seen by the neural network. The wavelength is $800$~nm, the peak laser intensity is $2.5\times10^{14}$ W/cm$^2$, and the laser pulse duration is $n_p=2$ cycles. (a) shows the distribution calculated for atomic nuclei moving classically. (b) displays the momentum distribution calculated within the quantum approach [Eq. (\ref{pmd_2})]. (c) shows the distribution calculated with semiclassical treatment of the nuclear motion [Eq.~(\ref{dist_res})].}
\label{pmds}
\end{figure}

It is well known that any neural network uses randomness in the training process, see, e.g., Ref.~\cite{Goodfellow2016}. We train $5$ different CNNs on the same training data sets (the set used in Ref.~\cite{Shvetsov2022} and the set we employ here for transfer learning) and we apply them to a test set of $N=100$ distributions obtained from the TDSE, Eq.~(\ref{tdse_t}), for different delays between the initial time and the probe pulse. An example distribution from the test set is shown in Figure~\ref{pmds}~(a). As in Refs.~\cite{Shvetsov2022,Shvetsov2023}, the peak laser intensities were chosen randomly between $1.0\times10^{14}$ and $4.0\times10^{14}$ W/cm$^2$. It is assumed that the probe pulse cannot start before the initial time $t=0$. Together with the non-negligible duration of the probe pulse, this implies a restriction on the minimum $R$ that we use [see Figure~\ref{calc_1}~(a) for an illustration]. For a two-cycle laser pulse this minimum internuclear distance is $3.67$~a.u.

The predictions averaged over 5 CNNs for every time delay are shown in Figure~\ref{calc_1}~(b). Assuming Gaussian distributed predictions, we estimate absolute errors of all the predicted internuclear distances. The overall agreement between the results obtained with ML and the actual time-dependent internuclear distance $R\left(t\right)$ is quite good except at the time delays $t \geq 300$~a.u., see Figure~\ref{calc_1}~(b). We attribute the discrepancies for large time delays to the fact that due to the finite pulse duration the PMDs contain contributions from $R>8.0$~a.u. Recall that the CNN of Refs.~\cite{Shvetsov2022,Shvetsov2023} was trained on a set of PMDs obtained for $1.0<R<8.0$~a.u. We find that for ionization by a two-cycle laser pulse, one CNN randomly selected from $5$ CNNs retrieves the bond length with an MAE of $0.26$~a.u. Simultaneously, application of a single-cycle probe pulse provides for this CNN a smaller MAE of $0.21$~a.u. Obviously, short probe pulses imply small displacements of the nuclei during the pulse, and, as a result, the retrieval of the bond length becomes more accurate.

\begin{figure}[h]
\centering
\includegraphics[width=.70\textwidth]{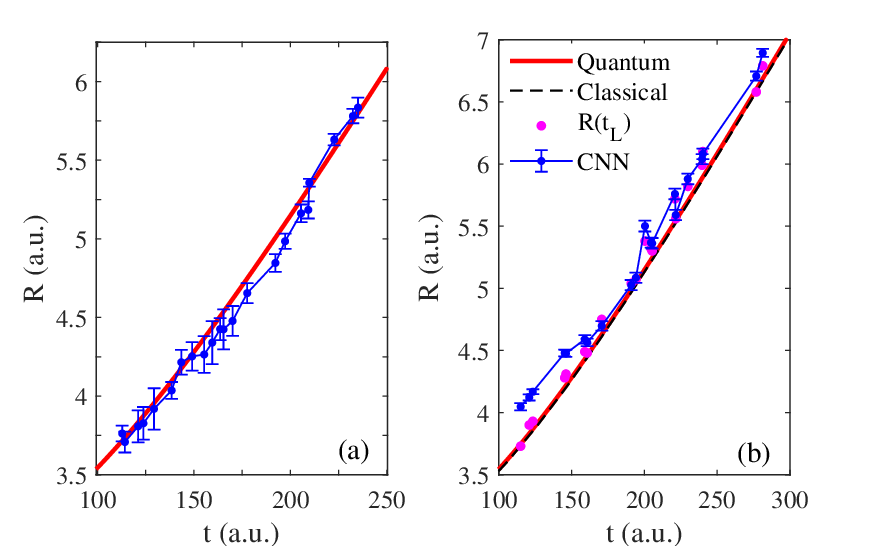}
\caption{(a) Plot of averaged predictions of 5 CNNs for the internuclear distances (blue points) at different time delays vs. the time-dependent expectation value of the internuclear distance (red curve) calculated from the solution of the TDSE~(\ref{tdse_nuc}). (b) Retrieval of the time-dependent bond length in the semiclassical approach. Blue points show average of predictions of $50$ CNNs for internuclear distances at various time delays. The internuclear distance obtained from Newton's equation of motion and the expectation value of the internuclear distance calculated from the TDSE (\ref{tdse_nuc}) are shown by the dashed black and the thick red curves, respectively. The magenta points correspond to the internuclear distances calculated from Eq.~(\ref{r_res}).}
\label{calc_2}
\end{figure}

In the second approach, we treat nuclear motion quantum-mechanically. More specifically, we solve the TDSE for the nuclear motion
\begin{equation}
i\frac{\partial}{\partial t}\phi\left(R,t\right)=\left[-\frac{1}{2M}\frac{\partial^2}{\partial R^2}+E_{2p\sigma_{\mu}}\left(R\right)\right]\phi\left(R,t\right),
\label{tdse_nuc}
\end{equation}
where $\phi\left(R,t\right)$ is the nuclear wave function, and $E_{2p\sigma_{\mu}}$ is the energy of the first excited state of the H$_2^+$ ion. Here we assume $\phi\left(R,t=0\right)=\phi_{1\sigma_{g}}$, i.e., instantaneous vertical excitation in the pump step. The PMD that corresponds to a given time delay $\Delta t$ is calculated as follows:
\begin{equation}
\frac{dP}{d^{3}\vec{k}}=\int\left|\phi\left(R,\Delta t\right)\right|^2\frac{dP_{\textrm{\footnotesize fr}}}{d^{3}\vec{k}}\left(R\right)dR,
\label{pmd_2}
\end{equation}
where the distribution $dP_{\textrm{\footnotesize fr}}/d^{3}\vec{k}\left(R\right)$ is obtained from the TDSE for frozen nuclei. Therefore, the PMD (\ref{pmd_2}) is an incoherent sum of the distributions for fixed internuclear distances which are weighted with the modulus square of the nuclear wave function at $t=\Delta t$. Figure \ref{pmds}~(b) shows an example of a PMD calculated from Eq.~(\ref{pmd_2}). As in the first case, we choose random peak laser intensities in the range between $1.0\times10^{14}$ and $4.0\times10^{14}$ W/cm$^2$. We calculate a test set of $N=100$ distributions (\ref{pmd_2}) and use them as an input for the 5 CNNs. The performance of ML on this set is illustrated by Figure~\ref{calc_2}~(a). The MAE of the retrieved $R$ for one arbitrarily chosen CNN is equal to $0.074$~a.u. The reduced range of the time delays in this calculation is due to the non-zero width of the spreading nuclei wave function. More specifically, we choose the range of the time delays so that the rising (decaying) edge of the nuclear wave function is negligibly small in the region $R<1.0$~a.u. $\left(R>8.0~\textrm{a.u.}\right)$ so that the integration range in Eq.~(\ref{pmd_2}) can be set to $\left[1,8\right]$. We note that this quantum approach does not account for the nuclear motion during the laser pulse. 
To include both the non-zero width of the nuclear wave packet and the nuclear motion during the pulse, we introduce the third, semiclassical approach based on a position-momentum quasiprobability distribution. More specifically, we use the Husimi distribution \cite{Husimi1940}, to sample the internuclear distances $R$ and velocities $V_{R}\equiv dR/dt$ of the initial wave packet. We calculate the Gabor transformation of the initial nuclear wave function $\phi\left(R,0\right)$:
\begin{equation}
G\left(R_{0},P_{R}\right)=\frac{1}{2\pi}\int\phi\left(R,0\right)\exp\left[-\frac{\left(R-R_{0}\right)^2}{2\delta^2}\right]\exp\left(-iP_{R}R\right)dR,
\label{gabor}
\end{equation}
where $P_R=MV_{R}$, and $\delta$ is the width of the exponential window. The squared modulus of the Gabor transformation (\ref{gabor}), i.e., $\left|G\left(R_{0},P_{R}\right)\right|^2$, is the Husimi distribution at time $t=0$. For a set of $N_{0}=10$ arbitrarily chosen initial bond lengths $R_{j}\left(t=0\right)$ and velocities $V_{R_j}\left(t=0\right)$, we solve Newton's equation of motion (\ref{Newton}) and calculate the classical internuclear distances $R_{j}\left(t\right)$ $\left(j=1,...,N_{0}\right)$. For every trajectory $R_{j}\left(t\right)$ we solve the electronic TDSE (\ref{tdse_t}) and calculate the corresponding electron momentum distribution. For a given time delay $\Delta t$, the total resulting PMD reads
\begin{equation}
\frac{dP\left(R\right)}{d^{3}\vec{k}}=\frac{1}{\sum_{j=1}^{N_0}w_{j}}\sum_{j=1}^{N_0}w_{j}\frac{dP_{j}}{d^{3}\vec{k}}\left(\Delta t\right),
\label{dist_res}
\end{equation}
where $w_{j}=\left|G\left(R_{j}\left(t=0\right),P_{R_j}\left(t=0\right)\right)\right|^2$ are used as weights. An example of a PMD calculated from Eq.~(\ref{dist_res}) is shown in Figure~\ref{pmds}~(c). The internuclear distance, which we assign as a label to the distribution (\ref{dist_res}), is given by
\begin{equation}
R=\frac{1}{\sum_{j=1}^{N_0}w_{j}}\sum_{j=1}^{N_0}w_{j}R_{j}\left(\Delta t\right).
\label{r_res}
\end{equation}
The semiclassical approach allows us to account for the nuclear motion during the laser pulse. Simultaneously, the ensemble of various trajectories $R_{j}\left(t\right)$ simulates the nuclear wave packet. In order to avoid computational costs, we fix the laser intensity to $2.52\times10^{14}$ W/cm$^2$ and consider a set of $20$ different time delays. We calculate the corresponding PMDs and use them to test $50$ different CNNs trained as outlined above. The averaged predictions of the CNNs are in good agreement with the bond length obtained from Eq.~(\ref{r_res}), except for some deviations at small and large time delays, see Figure~\ref{calc_2}~(b). It appears natural to explain these small discrepancies by the fact that the ionization yield depends on the nuclear trajectory, which is taken into account in Eq.~(\ref{dist_res}) for the momentum distribution but is ignored in Eq.~(\ref{r_res}). In order to test this assumption, we replace Eq.~(\ref{r_res}) by 
\begin{equation}
R=\frac{1}{\sum_{j=1}^{N_0}w_{j}Y_{\textrm{\footnotesize fin},j}}\sum_{j=1}^{N_0}w_{j}Y_{\textrm{\footnotesize fin},j}R_{j}\left(\Delta t\right),
\label{r_res2}
\end{equation}
where $Y_{\textrm{\footnotesize fin},j}$ is the total ionization yield corresponding to the internuclear distance $R_{j}\left(\Delta t\right)$. However, this modification only slightly affects the values of $R$. 

Therefore, the reasons for the deviations require further studies. Nevertheless, the results obtained here show that the neural network trained on the PMDs for fixed internuclear distances can be used to retrieve the time-dependent bond length with good accuracy. 

\section{Conclusions and outlook}

In conclusion, we have investigated the ability of deep learning to retrieve the time-dependent bond length in a dissociating molecule from electron momentum distributions produced by a strong laser field. To this end, we have simulated a pump-probe scheme, in which the pump step excites the molecule to the first excited electronic state, initiating nuclear motion, and the molecule is ionized by the probe pulse acting after a certain time delay. The corresponding PMDs have been calculated within the three different approaches. First, we have assumed that the nuclei move classically along the Born-Oppenheimer potential. In the second approach, the PMD corresponding to a given time delay was calculated as an incoherent sum of the distributions for different fixed internuclear distances weighted with the modulus square of the time-dependent nuclear wave function. Since this method does not include the nuclear motion during the laser pulse, we have also used the third, semiclassical approach, in which the initial conditions are sampled from the Husimi quasiprobability distribution. We have found that in all these three cases, the CNN trained on distributions obtained for fixed internuclear distances predicts the time-dependent bond length with a good accuracy. Therefore, our present results show that deep learning can be used not only for static, but also for dynamic molecular imaging based on electron momentum distributions.

The proposed approach can be straightforwardly extended to the three-dimensional case, and, by applying the transfer learning technique, to the case of focal volume averaged PMDs, see Ref.~\cite{Shvetsov2023}. Developments in these directions are already on the way. It will be interesting to apply ML for retrieval of not only the time-dependent internuclear distance, but also its velocity. It is also of interest to go beyond the the pump-probe scheme, i.e., to investigate whether the neural network is able to reconstruct the time evolution of the internuclear distance during the pulse from only one momentum distribution. Indeed, such a study will shed light on the ``memory" of a single momentum distribution with regard to the time-dependent information about the molecular ion, which is important for the development of efficient tools for time-resolved molecular imaging. The results obtained here suggest that deep learning is a powerful approach to this problem. 

\ack 
We are grateful to P.~Winter and Dr. S.~Yue for valuable discussions and continued interest to this work. This work was supported by the Deutsche Forschungsgemeinschaft (Grant No. SH 1145/1-2).

\end{document}